\documentclass[aps,pra,superscriptaddress,twocolumn,showpacs,nofootinbib]{revtex4-2}
\usepackage{silence}
\usepackage{amsmath,amssymb,amsfonts}
\usepackage{eqnarray}
\usepackage{graphics,float}
\usepackage{latexsym}
\usepackage{bm}
\usepackage{braket}
\usepackage{psfrag}
\usepackage{physics}
\usepackage{microtype}
\usepackage{lineno}
\usepackage{mathtools}
\usepackage{soul,color,xcolor,colortbl}
\usepackage{tikz}
\usepackage{lipsum}
\usepackage{comment}
\usepackage[title]{appendix}
\usepackage{pdfpages}
\usepackage{pgffor}

\renewcommand{\eqref}[1]{\unskip\,(\ref{#1})}

\usetikzlibrary{shapes}
\usepackage[colorlinks=true,
linkcolor=blue,
filecolor=magenta,
urlcolor=blue,
citecolor=blue]{hyperref}
\allowdisplaybreaks
\makeatletter
\AtBeginDocument{\let\LS@rot\@undefined}
\makeatother

\AtBeginDocument{%
	\setlength{\linenumbersep}{0.1cm}%
}

\newcommand{\marker}[2]{\raisebox{0.5pt}{\tikz{\node[draw,scale=0.65,#2,color=#1,fill=#1](){};}}}
\newcommand{\markertriangle}[1]{\raisebox{0.5pt}{\tikz{\node[draw,scale=0.4,regular polygon, regular polygon sides=3,color=#1,fill=#1](){};}}}
\newcommand{\markersquare}[2]{\raisebox{0.5pt}{\tikz{\node[draw,scale=0.8,#2,color=#1,fill=#1](){};}}}

\definecolor{lightblue}{rgb}{0.09,0.37,0.83}

\graphicspath{{Plots/}}
\begin{document}
\title{Induced supersolidity and hypersonic flow of a dipolar Bose-Einstein Condensate in a rotating bubble trap}
\author{Hari Sadhan Ghosh}
\email{harisadhanghosh4@gmail.com}
\affiliation{Department of Physics, Indian Institute of Technology Kharagpur, Kharagpur 721302, India}
\author{Soumyadeep Halder}
\email{soumya.hhs@gmail.com}
\affiliation{Department of Physics, Indian Institute of Technology Kharagpur, Kharagpur 721302, India}

\author{Subrata Das}
\email{subrata@vt.edu}
\affiliation{Department of Physics, Indian Institute of Technology Kharagpur, Kharagpur 721302, India}
\affiliation{Department of Physics, Virginia Tech, Blacksburg, Virginia 24061, USA}

\author{Sonjoy Majumder}
\email{sonjoym@phy.iitkgp.ac.in}
\affiliation{Department of Physics, Indian Institute of Technology Kharagpur, Kharagpur 721302, India}

\setstcolor{red}
\date{\today}
\begin{abstract}
    Motivated by the recent realization of space-borne Bose-Einstein Condensates (BECs) under micro-gravity conditions, we extend the understanding of ultracold dipolar bosonic gases by exploring their behavior in a novel trapping configuration known as the ``bubble trap" topology. Utilizing the three-dimensional numerical simulations within the extended Gross-Pitaevskii framework, we unveil diverse ground state phases in this static curved topology. Subsequently, we investigate the influence of rotation on a dipolar BEC confined to the surface of a spherical bubble. Our findings reveal that the rotation of a bubble trap at certain rotation frequencies can modify the effective local dipole-dipole interaction strength, leading to the induction of supersolidity and the formation of quantum droplets. In addition, we demonstrate that a bubble trap can sustain high circulation, with the flow persisting for a longer time. Significantly, adjusting the rf detuning parameter allows the condensate to achieve hypersonic velocity. Finally, we also explore the impact of drastic change in the topological nature of the trap on the rotating dipolar BEC, transitioning from a filled shell trap to a bubble trap and vice versa. Based on the results of the topological transition, we propose a dynamic protocol to drive the interacting atomic gas into the quantum Hall regime.
\end{abstract}

\maketitle

\section{Introduction}
The exploration of trapped atomic Bose-Einstein Condensates (BECs) has traditionally been steered by the investigations into atom-atom interaction strength, dimensionality, and the topological nature of the trapped geometry. The ability to tune the interaction strength between the atoms of ultracold atomic gases \cite{kwon_2021_spontaneous, marinescu_1998_atom, roberts_1998_resonant, theis_2004_tuning} gives rise to a plethora of fascinating physical phenomena, including the superfluidity \cite{leggett_1999_superfluidity}, formation of quantized vortices \cite{roccuzzo_2020_rotating, gallemi_2020_quantized, klaus_2022_observation, madison_2000_vortex, aboshaeer_2001_observation}, solitons \cite{burger_1999_dark, khaykovich_2002_formation, strecker_2002_formation, cornish_2006_formation, raghun_2015_two_dimensional}, the superfluid to Mott insulator phase transition in an optical lattice \cite{greiner_2002_quantum, colussi_2023_lattice}, and more recently, the formation of quantum droplets \cite{baillie_2018_droplet,baillie_2016_self, ferrierbarbut_2016_observation, wachtler_2016_ground} and the supersolid phase \cite{chomaz_2019_long, bottcher_2019_transient, tanzi_2019_observation}. The signature of supersolidity was initially detected in two experiments: one utilizing BECs within an optical cavity \cite{leonard_2017_supersolid}, and another with mixtures featuring spin-orbit coupling \cite{li_2017_a_stripe}. In recent years, the realization of quantum droplets and supersolid states in dipolar bosonic gases of rare-earth Dy \cite{kadau_2016_observing, wachtler_2016_ground, ferrierbarbut_2016_observation} and Er \cite{chomaz_2016_quantum,petter_2019_probing,chomaz_2019_long} atoms has attracted significant attention. Theoretical and experimental studies over the past few years have revealed the appearance of roton excitations \cite{blakie_2012_roton, chomaz_2018_observation, petter_2019_probing, natale_2019_excitation, hertkorn_2021_supersolidity, schmidt_2021_roton, bisset_2013_roton, santos_2003_roton, kirkby_2023_spin}, the formation of the self-bound quantum droplets \cite{ferrierbarbut_2016_observation, baillie_2016_self, baillie_2017_collective, schmitt_2016_self}, and the supersolid state in quasi one \cite{smith_2023_supersolidity, blakie_2020_supersolidity, tanzi_2019_observation, poli_2021_maintaining, roccuzzo_2019_supersolid, bland_2022_alternating, scheiermann_2023_catalyzation} and two-dimensional \cite{hertkorn_2021_supersolidity, norcia_2021_two_dimensional, bl_2022_two_dimensional, halder_2022_control, halder_2023_two_dimensional,halder_2024_induced} harmonically trapped confinements. \par

The trap geometry of a BEC plays a crucial role in shaping various intriguing physical phenomena. Examining BECs in one dimension (1D) provides a deeper understanding of fermionization \cite{kinoshita_2004_observation} and many-body systems in non-equilibrium conditions \cite{hofferberth_2007_non_equilibrium}. In two dimensions (2D), the study of BECs has shed light on quasi-condensation, the Berezinskii-Kosterlitz-Thouless (BKT) transition \cite{filinov_2010_berezinskii, bombin_2019_berezinskii} and the formation of a two-dimensional vortex lattice \cite{zhang_2005_vortex}. Condensates in double-well and optical lattice potential have numerous applications, such as matter-wave interferometry \cite{shin_2004_atom, schumm_2005_matter} and spin squeezing \cite{esteve_2008_squeezing}. Interestingly, a rotating toroidal or ring trap can maintain a continuous superfluid flow with substantially high angular momentum, which persists for a longer time \cite{ramanathan_2011_superflow, ryu_2007_observation, moulder_2012_quantized, murray_2013_probing, corman_2014_quench}. Such a high angular momentum state has an analogy with the quantum system of charged particles in a uniform magnetic field, relevant for condensed matter physics such as type II superconductors or the quantum Hall effect \cite{cooper_2008_rapidly, bloch_2012_quantum}. However, generating a state with such high angular momentum in harmonic trap requires the trap's rotation frequency to be close to the radial trapping frequency, which can lead to the loss of the atoms from the condensate. In contrast, a higher angular momentum state can be easily generated by rotating a ring or toroidal trap. Theoretical studies predict that this higher angular momentum state in ring trap could provide a path to realize quantum hall state \cite{roncaglia_2011_rotating}.\par
Recent advancements in the precision and sophistication of controlling the trapping mechanisms for ultracold atomic gases have led to the formation of a novel trapping geometry resembling a bubble, where all the atoms in the condensate are confined to the surface of the bubble \cite{carollo_2022_observation,zobay_2004_atom,  tononi_2019_bose}. The unique topology and curved geometry of this bubble trap opens a promising new avenue for intriguing research endeavors, including the investigations into the influence of curvature on various topological defects and their dynamics \cite{tononi_2022_topological, bereta_2021_superfluid, tomishiyo_2024_superfluid, white_2024_triangular}, exploration of new collective modes \cite{lannert_2007_dynamics, sun_2018_static, padavi_2018_physics, diniz_2019_ground}, examination of matter-wave interference \cite{lannert_2007_dynamics}, comprehension of the persistent flow of superfluid \cite{guo_2020_supersonic}, the emergence of the supersolid phase in soft-core BECs \cite{ciardi_2024_supersolid, santi_2019_softcore}, and the study of dimensional crossover from harmonic to thin bubble trap configurations \cite{sun_2018_static, rhyno_2021_thermodynamics, tononi_2023_low_dimensional}. Recent studies have also unveiled the impact of rotation and gravity on bubble-trapped superfluid \cite{li_2023_equatorial, saito_2023_rossby, arazo_2021_gravity}.\par

Motivated by this optimistic trapping configuration and the emergence of quantum droplets and supersolid phases in dipolar BECs, in this article, we study the dipolar BEC confined on the surface of a bubble trap and the influence of the trap rotation on it. Remarkably, we find that the rotation of the bubble trap with certain rotation frequencies alters the local relative dipole-dipole interaction (DDI) strength, which could induce supersolidity and the nucleation of quantum droplets in an otherwise superfluid state. Additionally, we note the ability to sustain a state with exceptionally high circulation within this bubble-shaped topology. This flow with high angular momentum per particle also persists for a longer duration. Moreover, we observe that a rotating bubble trap, characterized by a large radius and strong transverse confinement, leads to the attainment of hypersonic velocity of the condensate. Furthermore, we have investigated the impact of altering the topology of the trap, transitioning from a filled spherical shell to a thin shell and vice versa on the dipolar BEC.\par
The paper is organized as follows. In Section \ref{sec2}, we present the extended Gross-Pitaevskii theory, discuss the bubble trap potential, and outline the procedure to calculate superfluid fraction. We provide the non-rotating ground state phases depending on the s-wave scattering lengths in Sec. \ref{sec3}. The impact of the trap's rotation on the superfluid is discussed in Sec. \ref{sec4}, where we observe rotation-induced supersolidity, and analyze the modification of DDI energy and the gain of the angular momentum of the system. In Sec. \ref{sec5}, we examine the effect of topological transition on the condensate by tuning the trap parameters and propose a potential way to realize quantum hall states in a harmonic trap. Finally, we draw important conclusions and provide future perspectives in Sec. \ref{sec6}. In Appendix \ref{appA1}, we explore the effect of rotation on the supersolid phase in detail. The stability of the hypersonic flow is discussed in Appendix \ref{appA2}.

\section{Formalism}

At zero kelvin temperature, the dipolar BEC of $N$ atoms with mass $m$ and magnetic dipole moment $\mu$ is described by the macroscopic wave function $\psi(\mathbf{r},t)$, whose temporal evolution is governed by extended Gross-Pitaevskii equation (eGPE) $i\hbar\dot{\psi}(\mathbf{r},t)=\hat{H}\psi(\mathbf{r},t)$ \cite{wachtler_2016_quantum,chomaz_2016_quantum,bisset_2016_ground,lima_2011_quantum} with the Hamiltonian,

\begin{align}
     & \hat{H}=\Big[-\frac{\hbar^2}{2m}\nabla^2 +V(\mathbf{r})+ g|\psi(\mathbf{r},t)|^2 \nonumber                                          \\
     & +\int U_{\rm dd}(\mathbf{r}-\mathbf{r'})|\psi(\mathbf{r'},t)|^2 d\mathbf{r'} +\gamma(\epsilon_{\rm dd})|\psi(\mathbf{r},t)|^3\Big].
    \label{gpe}
\end{align}

\begin{figure}[hb]

    \includegraphics[width=0.4\textwidth]{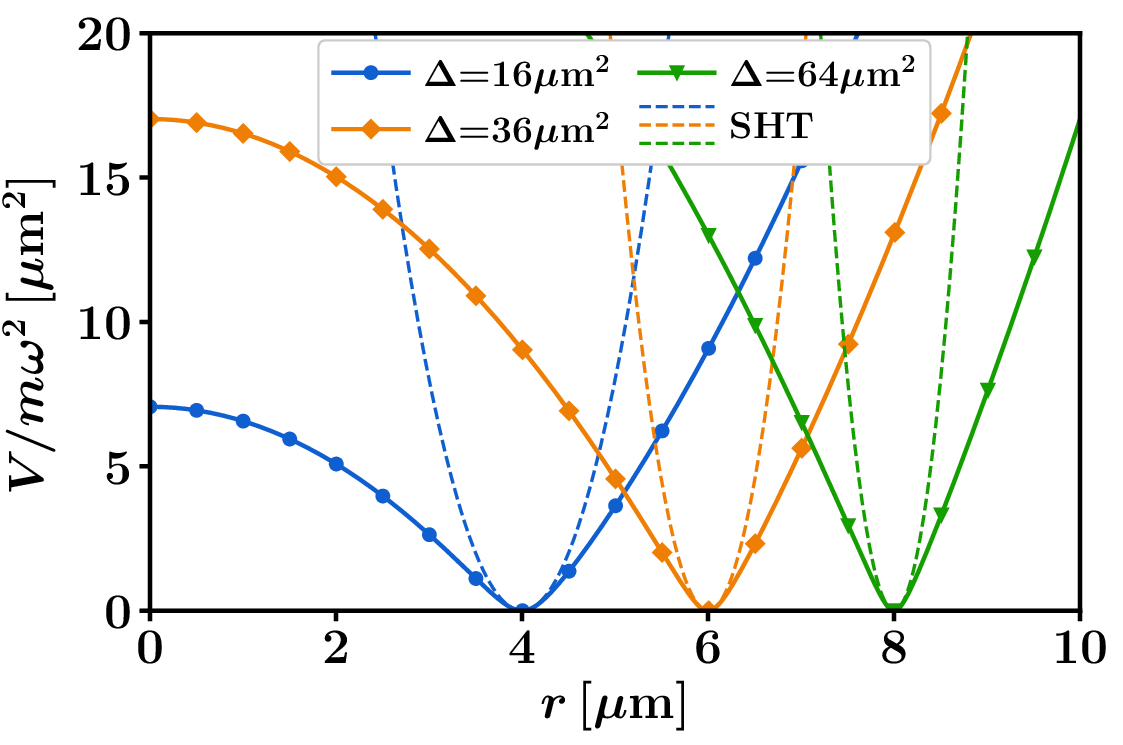}
    \caption{Comparison between bubble trap potentials as defined by Eq. (\ref{bubble}) with $\Theta=1~\mu \rm{m}^2$ for different values of $\Delta$. The dashed line plots of respective color demonstrate their closely approximated radially shifted harmonic trap (SHT) potential with the same $\Delta$ values.}
    \label{fig:fig1}

\end{figure}

\noindent Where $\psi(\mathbf{r},t)$ is normalized with $\int|\psi(\mathbf{r},t)|^2d\mathbf{r}=N$. The contact interaction strength $g=4\pi \hbar^2a_s/m$ is fixed by the s-wave scattering length $a_s$. The DDI potential is described by the term $U_{\rm dd}(\mathbf{r})=\frac{\mu_0\mu^2}{4\pi}\frac{1-3(\hat{\mathbf{e}}_z.\hat{\mathbf{r}})^2}{r^3}$, where $\mu_0$ is the permeability of vacuum, $\hat{\mathbf{e}}_z$ is the unit vector in the $z$-direction, and $\mathbf{r}$ is the separation vector between the dipoles. Here, the dipoles are polarised in the $z$-direction. The last term in Eq. (\ref{gpe}) takes into account the quantum fluctuations \cite{petrov_2015_quantum} in terms of the first order beyond mean-field Lee-Huang-Yang correction, with $\gamma(\epsilon_{\rm dd})=(128\sqrt{\pi}\hbar^2a_s^{5/2}/3m)(1+\frac{3}{2}\epsilon_{\rm dd}^2)$ \cite{lima_2012_beyond, lima_2011_quantum, fischer_2006_mean}. Here the dimensionless parameter $\epsilon_{\rm dd}=a_{\rm dd}/a_s$ defines the ratio between the strength of DDI and contact interaction, where $a_{\rm dd}=m\mu_0\mu^2/12\pi\hbar^2$ is the dipolar length. $V(\mathbf{r})$ is the bubble trap \cite{sun_2018_static, zobay_2001_two_dimensional, zobay_2004_atom} potential with trap frequency $\omega$, and it is modelled as
\begin{equation}
    V(\mathbf{r})=m\omega^2\sqrt{(r^2-\Delta)^2/4+\Theta^2}-m\omega^2\Theta. \label{bubble}
\end{equation}
Here $\Delta$ represents the detuning between the applied rf field and the participating energy states of the condensate, and $\Theta$ denotes the Rabi frequency between these states. These parameters control the radius and width of the bubble trap. For $\Delta=\Theta=0$, the bubble trap (Eq. (\ref{bubble})) reduces to a well-known spherical harmonic trap. For a fixed value of $\Theta$ and with the larger value of $\Delta (\Delta>\Theta)$, we get a thinner spherical shell potential with a larger radius which can be approximated as a radially shifted harmonic trap (SHT) $V_{\rm SHT}(\mathbf{r})=\frac{m\omega^2\Delta}{2\Theta}(r-\sqrt{\Delta})^2$ \cite{sun_2018_static}, as shown in Fig. \ref{fig:fig1}.\par
\subsection{Superfluid Fraction}
A dipolar BEC can exhibit superfluid, supersolid, or droplet phases depending on the value of $\epsilon_{\rm dd}$. The various ground states are distinguished by their superfluid fraction. It is defined through the non-classical rotational inertia according to $f_s^{\Omega}=1-I/I_{\rm {cl}}$ \cite{leggett_1970_can_a, tanzi_2021_evidence}, where $I=\lim_{\Omega\to 0}\langle\hat{L}_z\rangle/\Omega$ and $I_{\rm{cl}}=\int d\mathbf{r}(x^2+y^2)|\psi(\mathbf{r},t)|^2$ is its classical rigid body moment of inertia in the $z=0$ plane. It is measured by applying a perturbation $-\Omega\hat{L}_z$ to the system, where $\Omega\rightarrow 0$ is the rotation frequency of the trap about the $z$-axis and $\hat{L}_z$ is the $z$-component of the angular momentum operator. In the case of an axially symmetric condensate, the non-classical rotational inertia $f_s^{\Omega}$ matches with the superfluid fraction $f_s$ \cite{leggett_1998_superfluid, leggett_1970_can_a}. In our case, the superfluid fraction is estimated by Leggett's upper bound \cite{leggett_1998_superfluid}
\begin{align}
    f_s^{\sigma} & =\Bigg[\expval{n(\sigma)}\expval{\frac{1}{n(\sigma)}}\Bigg]^{-1}\nonumber                       \\
                 & = (2L)^2\Bigg[\int_{-L}^{L}d\sigma ~n(\sigma)\int_{-L}^{L}\frac{d\sigma}{n(\sigma)}\Bigg]^{-1}.
    \label{leget}
\end{align}
Here, $\sigma={x,y} \in[-L,L]$ and $n(\sigma)=\int\int dydz|\psi(x,y,z)|^2$. The overall superfluid fraction in the $x$-$y$ plane is determined by taking an average of the superfluid fraction along $x$ and $y$-direction. We find that in our case $f_s{^x}=f_s{^y}=f_s$. In our analysis, we consider $L=6~\mu\rm{m}$ and characterize the state with $f_s\geq 0.9$ as the superfluid phase, whereas the supersolid phase and droplet phase corresponds to $0.1<f_s<0.9$ and $f_s\leq$0.1, respectively.
\label{sec2}
\section{Non-rotating ground state}
To begin our exploration of the rotating state in a bubble trap, it is imperative to first delve into the characteristics of the non-rotating states. We examine the ground state configurations by numerically evolving the eGPE in the imaginary time using the split-step Crank-Nicolson method \cite{crank_1947_practical}. For definiteness, we consider $N=6\times 10^4$ number of $^{164}$Dy atoms confined in a bubble trap with trap frequency $\omega=2\pi\times100$ Hz in all three directions, with the other parameters $\Delta=36l_{\rm{osc}}^2=21.9~\mu\rm{m}^2$ and $\Theta=0.0441l_{\rm{\rm{osc}}}^2$, where $l_{\rm{osc}}=\sqrt{\hbar/m\omega}=0.78~\mu \rm{m}$ is the characteristic length scale set by the trap. The $^{164}$Dy atom has the magnetic dipole moment $\mu=9.93\mu_{\rm B}$ and the corresponding dipolar length $a_{\rm{dd}}=131a_{\rm B}$, where $\mu_{\rm B}$ and $a_{\rm B}$ is the Bohr magneton and Bohr radius, respectively.\par
\begin{figure}[t]
    \centering
    \includegraphics[width=0.48\textwidth]{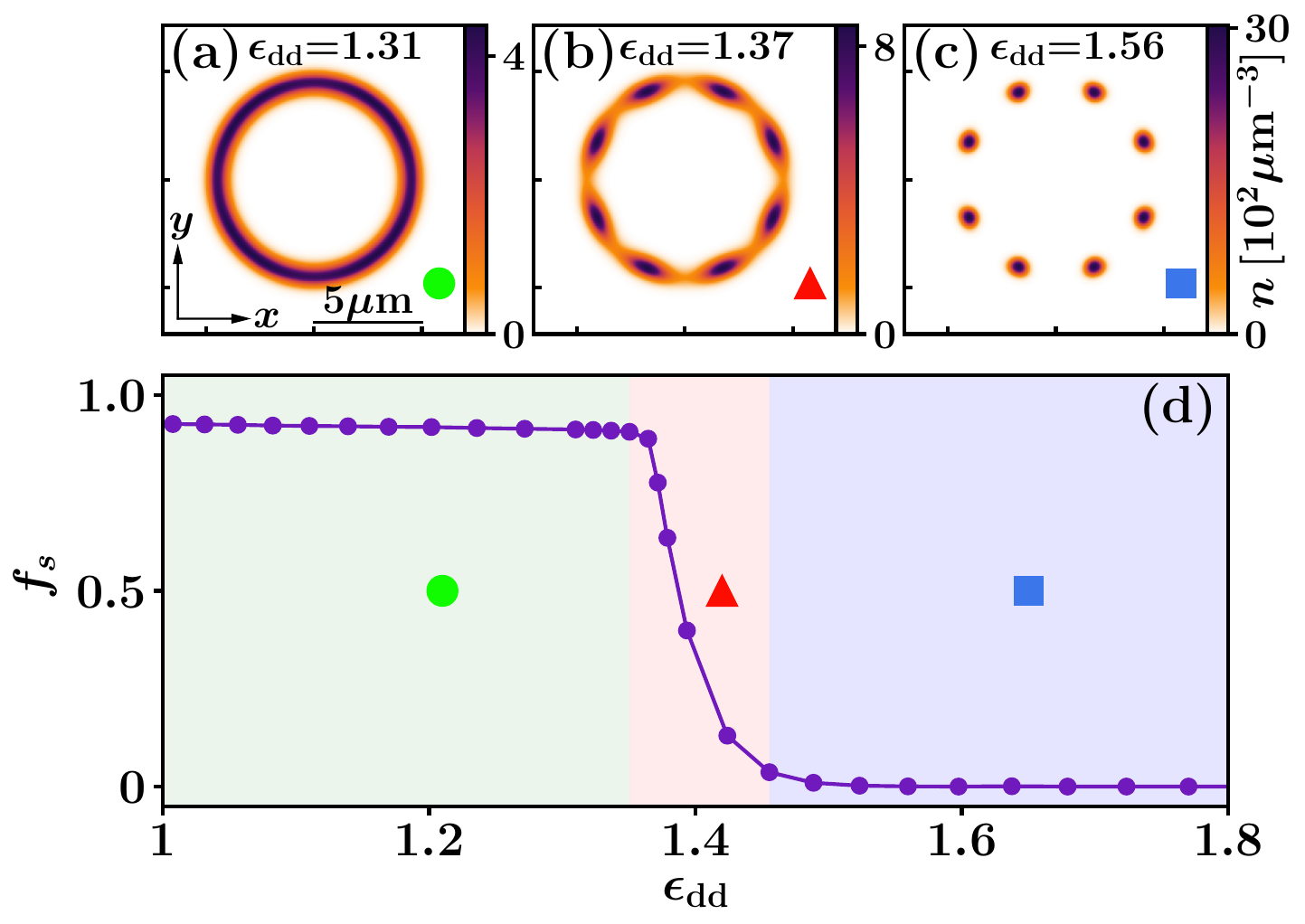}
    \caption{Non-rotating ground state density profiles in the $z=0$ plane (a)-(c) for different values of $\epsilon_{\rm dd}$ representing (a) superfluid (\protect\marker{green}{circle}), (b) supersolid (\protect\markertriangle{red}) and (c) incoherent crystal of isolated droplet (\protect\markersquare{lightblue}{rectangle}) phase in a bubble trap with $\Delta=21.9~\mu\rm{m}^2$ and $\Theta=0.0441l_{\rm{\rm{osc}}}^2$. (d) Shows the superfluid fraction $f_s$ as a function of dimensionless parameter $\epsilon_{\rm{dd}}$. The markers in (a)-(d) and the background hue in (d) signify the distinct phases.}
    \label{fig:fig2}
\end{figure}

The different ground state configurations are emerging based on the value $\epsilon_{\rm {dd}}$, which can be scaled experimentally by tuning the s-wave scattering length through the Feshbach resonance technique \cite{chin_2010_feshbach}. We obtain that the associated solutions for $\epsilon_{\rm dd}<1.35$ are fully superfluid BECs with unmodulated density distribution [see Fig. \ref{fig:fig2}(a)] having superfluid fraction close to unity ($f_s \geq 0.9)$ [see Fig. \ref{fig:fig2}(d) \protect\marker{green}{circle}]. As $\epsilon_{\rm dd}$ is increased to favor the DDI strength over contact interaction, beyond a critical value ($\epsilon_{\rm dd}\approx1.35$), a spontaneous density modulation is developed in the condensate due to the softening of the roton mode \cite{chomaz_2018_observation}. In the interval $1.35<\epsilon_{\rm dd}<1.43$, the ground state displays a modulated density profile [see Fig. \ref{fig:fig2}(b)] and it appears to be a regularly arranged ellipsoidal droplets along the trap, and each droplet is elongated along the polarisation direction and immersed in a background of low-density condensate with the superfluid fraction in the range of ($0.1<f_s<0.9$) [see Fig. \ref{fig:fig2}(d) \protect\markertriangle{red}]. With further increase in the value of $\epsilon_{\rm dd}$, the overlap between the droplets vanishes, and the ground state exhibits incoherent crystal of isolated droplets [see Fig. \ref{fig:fig2}(c)] with $f_s\leq0.1$ [see Fig. \ref{fig:fig2}(d) \protect\markersquare{lightblue}{rectangle}].\par
Depending on the value of $\epsilon_{\rm dd}$, the ground state in a bubble trap exhibits similar phases as dipolar BEC in a harmonic trap \cite{chomaz_2019_long, halder_2022_control}, albeit with distinct density distributions. The difference in the density distributions results from the bubble trap confinement, which leads to an exceptional arrangement of dipoles within the system. The dipoles are aligned in a head-to-tail fashion along the polar angle $\theta$, i.e., $\hat{\mathbf{e}}_z.\hat{\mathbf{r}}=1$, resulting in an attractive DDI. While, along the azimuthal angle $\phi$, where $\hat{\mathbf{e}}_z.\hat{\mathbf{r}}=0$, the dipoles are positioned side-by-side and interact repulsively. This specific arrangement yields distinct interactions for dipoles positioned along the equator and those at the pole. Equatorial dipoles experience a combination of attractive and repulsive forces, originating from their neighboring dipoles along the equator and those positioned above and below them along the meridian lines, respectively. Conversely, dipoles located at the poles encounter solely repulsive forces from their surrounding dipoles as a result of side-by-side arrangement. Consequently, the droplets adopt a curved shape along the surface of the bubble, exhibiting an ellipsoidal distribution elongated along the polarization direction (See three-dimensional density isosurfaces in Fig. S1 in Supplementary Material), and the density at the pole remains lower to minimize the overall energy of the system.\par
\label{sec3}
\section{Rotating ground state}
Once the ground state phases are established, we investigate the impact of rotation on the ground state configurations at different rotation frequencies within the superfluid phase regime (i.e., $\epsilon_{\rm dd} < 1.35$). We determine the rotating ground states by incorporating the angular momentum constraint $-\Omega \hat{L}_z$ into the Hamiltonian (Eq. (\ref{gpe})) and evaluating the modified eGPE in imaginary time. Here, $\hat{L}_z$ represents the $z$-component of the angular momentum operator, and the condensate is rotating with angular frequency $\Omega$ about the $z$-axis.\par
\begin{figure}[t]

    \includegraphics[width=0.48\textwidth]{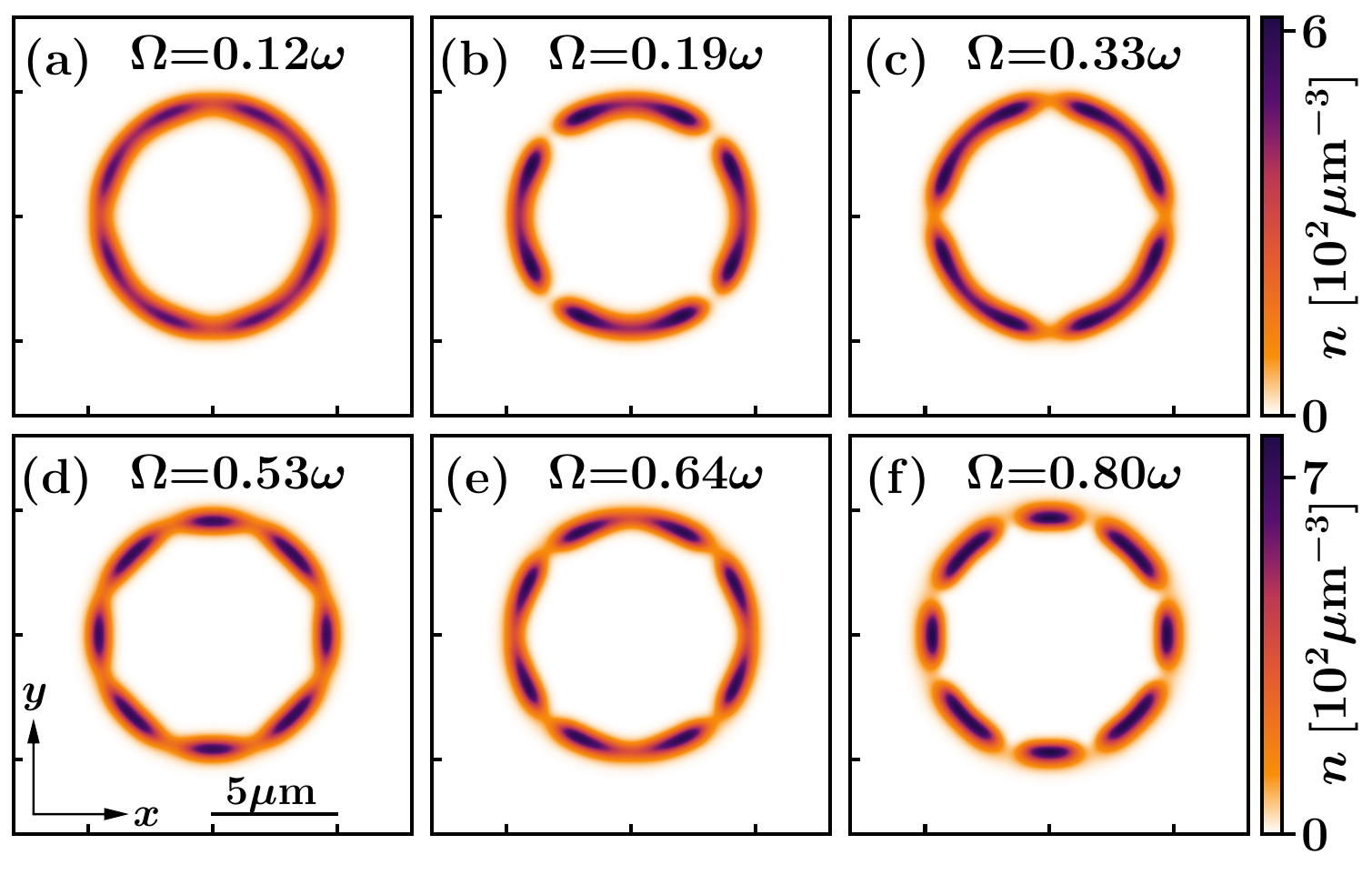}
    \caption{Rotation-induced spontaneous density modulation and formation of supersolid-like density distribution at specific ranges of rotation frequencies. (a)-(f) Show the rotating ground state density profiles with $\epsilon_{\rm{dd}}=1.31$ for $\Omega=0.12\omega, 0.19\omega,0.33\omega,0.53\omega,0.64\omega,$ and $0.80\omega$ in the $z=0$ plane, respectively. See Fig. S2 in the Supplementary Material for three-dimensional isosurface plots. Other parameters are same as of Fig. \ref{fig:fig2}(a).}
    \label{fig:fig3}
\end{figure}
\begin{figure}[t]

    \includegraphics[width=0.48\textwidth]{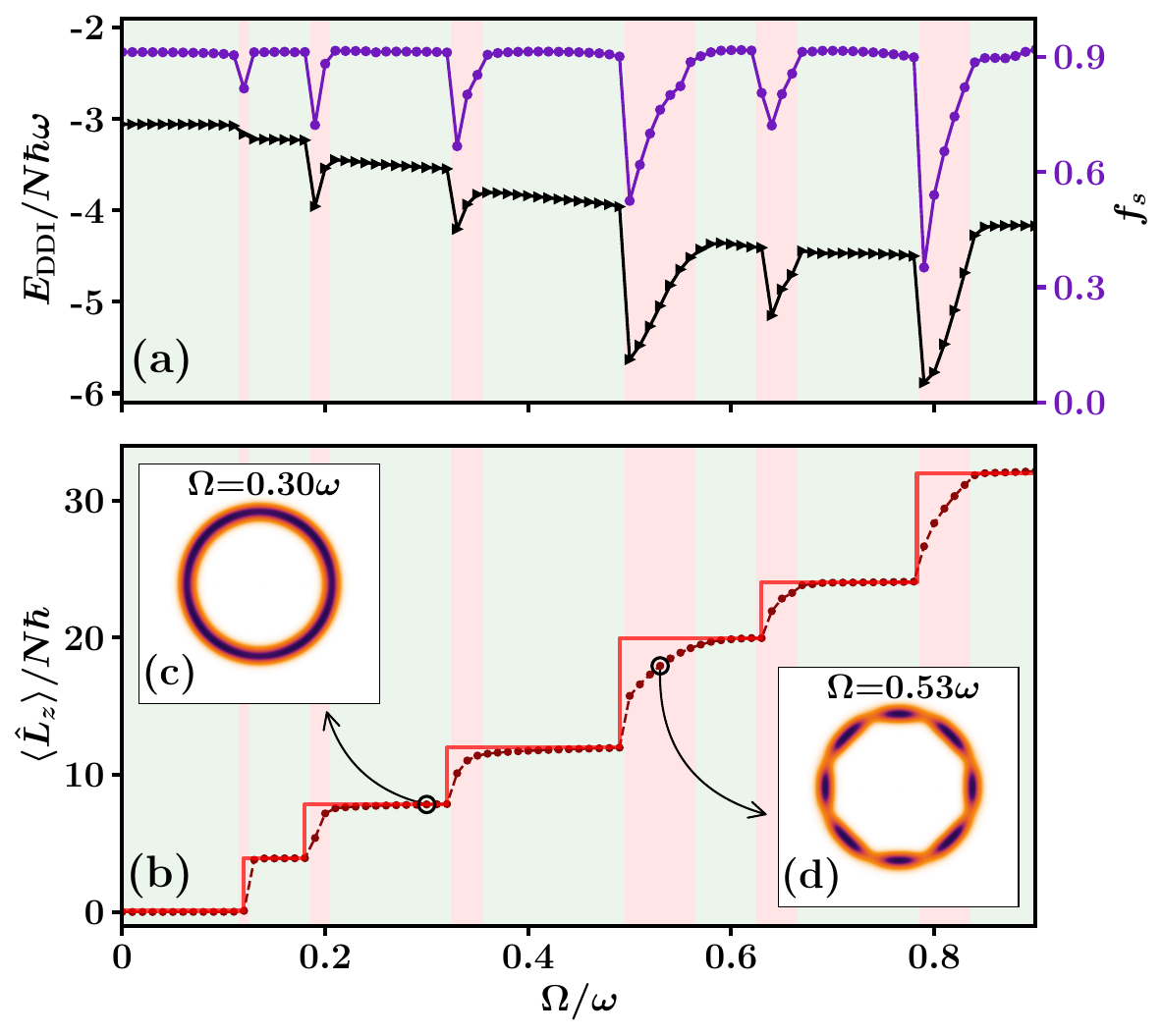}
    \caption{(a) Variation of DDI energy $E_{\rm{DDI}}/N\hbar\omega$ (black line with triangular markers) and superfluid fraction $f_s$ (purple line with circular markers) for rotating ground states with trap rotation frequency $\Omega$. The dips at specific ranges of $\Omega$ values resemble the increase in the attractive DDI which leads to the rotation-induced supersolid transition. (b) Shows angular momentum $\langle\hat{L}_z\rangle/N\hbar$ of the rotating ground state as a function of the rotation frequency $\Omega$ of the trap for $\epsilon_{\rm dd}=1.31$. The step-like growth (the red solid line is drawn for clarity) signifies the quantization of angular momentum corresponding to the superfluid phase, while the intermediate linear growth indicates a solid body response and characterizes the supersolid state. The insets (c) and (d) show the density distribution in the $z=0$ plane for $\epsilon_{\rm dd}=1.31$ at $\Omega=0.30\omega$ and $\Omega=0.53\omega$, respectively. The light green and the light red background specify the regions of the superfluid and supersolid state, respectively. Other parameters are same as of Fig. \ref{fig:fig2}(a).}
    \label{fig:fig4}
\end{figure}
For instance, at $\epsilon_{\rm dd}=1.31$ the non-rotating ground state exhibits a superfluid state with a smooth, unmodulated density distribution on the surface of the bubble as illustrated in Fig. \ref{fig:fig2}(a). However, as we introduce rotation at certain rotation frequencies, the condensate spontaneously develops density modulation and the system manifests a supersolid-like density configuration with periodic density humps on the curved surface of the bubble interconnected by lower density region [see Fig. \ref{fig:fig3}]. At these rotation frequencies, the DDI energy ($E_{\rm{DDI}}$) not only remains attractive in nature but also exhibits sudden dips in its value, as shown in Fig. \ref{fig:fig4}(a). These dips arise from the modification of the effective local DDI due to rotation at specific ranges of $\Omega$ values, emphasizing the pronounced role of DDI. As a consequence, the superfluid fraction of the condensate experiences an abrupt decrease at these specific ranges of $\Omega$ values, indicating the emergence of a supersolid state with $0.3<f_s<0.9$ [see Fig. \ref{fig:fig4}(a)]. The discontinuous change in the value $f_s$ indicates the first-order phase transition. On the other side, by tuning the scattering length, the transition from superfluid to supersolid represents a second-order phase transition characterized by a gradual change in $f_s$ [see Fig. \ref{fig:fig2}(d)]. A larger dip in the value of $E_{\rm{DDI}}$ and $f_s$ corresponds to a more significant modulation in density. However, this distinctive feature is not observed in either a rotating dipolar BEC in a harmonic trap or a non-dipolar BEC in a rotating bubble trap. The exclusive arrangement of dipoles due to bubble trap confinement, coupled with anisotropic DDI, makes the rotation-induced supersolid transition energetically preferable compared to the conventional formation of quantized vortices for these specific ranges of $\Omega$ values, as indicated by the sudden drop in DDI energy shown in Fig. \ref{fig:fig4}(a). As the scattering length decreases and, consequently, $\epsilon_{\rm dd}$ increases, the influence of the dipolar interaction becomes even more pronounced, leading to the reduction of superfluid density. So, in the supersolid regime, the value of the system's superfluid fraction drops from its initial value at specific rotation frequencies, approaching to the droplet regime (see Appendix \ref{appA1} for details). On the contrary with decreasing $\epsilon_{\rm dd}$, the modulation in the density profile becomes less prominent, and beyond a critical value ($\epsilon_{\rm dd}\approx 1$), no modulation occurs with rotation.\par

The angular momentum of the system also changes distinctively. In Fig. \ref{fig:fig4}(b), we depict the variation of the expectation value of angular momentum $\langle\hat{L}_z\rangle$ per particle of the rotating ground states as a function of the rotation frequency $\Omega$. Generally, trap rotation below a critical rotation frequency can not affect the superfluid state trapped in a harmonic trap. However, when the rotation frequency exceeds the critical value required for the vortex nucleation, the angular momentum exhibits a sudden jump from its initial zero value, reflecting the formation of quantized charged vortex \cite{madison_2000_vortex, aboshaeer_2001_observation, kawaguchi_2004_splitting, isoshima_2007_spontaneous}. The quantization of circulation restricts the value of the angular momentum to be fixed until the
rotation frequency is enough for nucleating further vortices. When the rotation frequency becomes sufficient, then again, the new vortices appear in the condensate, and the angular momentum again experiences a discrete jump in its value. Similarly, in a bubble trap, at certain rotation frequencies, new sets of vortex lines attempt to enter the system. However, the curved geometry of the trap and the tight confinement of the condensate on the bubble’s surface prevents these vortex lines from fully penetrating the condensate. Instead, they traverse close to the surface, causing density modulation in the system at these specific ranges of $\Omega$. Similar observations have been discussed in \cite{mukherjee_2022_crystallization,sinha_2005_two_dimensional}, where fast rotation and tight transverse confinement in an axially elongated trap lead to the arrangement of vortices along the major axis of the trap, resulting in the crystallization of the system. In our scenario, vortices enter along radial directions from the outer region, and when the vortex lines are close to the surface of the bubble, density gets modulated along with the linear increase in the angular momentum. As rotation frequency $\Omega$ is increased, these vortices migrate toward the center, producing an unmodulated superfluid state. With further increases in the rotation frequency $\Omega$, new vortex lines try to enter again, causing the condensate to undergo density modulation. As a result, we observe an alternating superfluid-supersolid phase domain as a function of $\Omega$, leading to the linear growth of angular momentum along with step-like growth [see Fig. \ref{fig:fig4}(b)].\par

Remarkably, we have found that the transfer of angular momentum is more efficient in the bubble trap confinement as compared to the harmonic trap, leading to higher angular momentum per particle at lower $\Omega$ values, with vortices nestled inside zero-density regions. As a result, the condensate circulates with significantly high angular momentum around the rotation axis. Notably, this flow persists even after halting the rotation of the trap as shown in Fig. S3 in Supplementary Material.\par

\begin{figure}[t]
    \centering
    \includegraphics[width=0.49\textwidth]{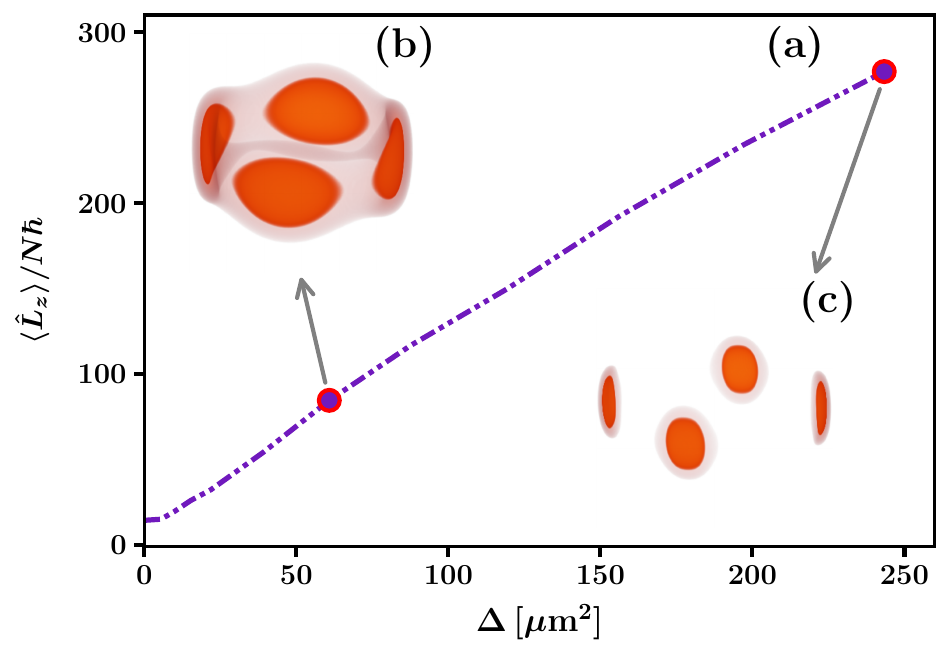}
    \caption{(a) Angular momentum $\langle\hat{L}_z\rangle/N\hbar$ of the ground state in the rotating frame shows linear variation with $\Delta$ for $\epsilon_{\rm dd}=1.31, \Omega=0.9\omega$ and $\Theta=0.0441l_{\rm{osc}}^2$. With these parameters, the inset (b) shows the density isosurfaces of a supersolid state at $\Delta=60.84~ \mu \rm{m}^2$, and the inset (c) shows the density isosurfaces of a droplet state formed by macrodroplets at $\Delta=243.4~ \mu \rm{m}^2$. Note that $\langle\hat{L}_z\rangle/N$ is approximate to $14\hbar$ for $\Delta=0$ and reaches close to $276\hbar$ at $\Delta=243.4 ~\mu \rm{m^2}$.}
    \label{fig:fig5}
\end{figure}

Moreover, there is a growing interest in preparing strongly correlated ultracold quantum bosonic gases within the quantum Hall regime \cite{bloch_2008_many, fetter_2009_rotating, ho_2001_bose}, but achieving the necessary angular momentum poses a challenge. The rotation frequency required to generate such substantial angular momentum in a harmonically trapped BEC surpasses the deconfinement frequency. However, the ability to generate larger angular momentum in the bubble trap than in the harmonic trap makes it a promising candidate for obtaining a quantum Hall state. Furthermore, in experiments, the $\Delta$ parameter can be tuned by modulating the rf detuning \cite{zobay_2004_atom}. Increasing the value of $\Delta$ for a fixed rotation frequency $\Omega$ results in a larger moment of inertia of the condensate and facilitates access to higher angular momentum. Fig. \ref{fig:fig5}(a), illustrates the growth of angular momentum with $\Delta$ for a constant rotation frequency $\Omega=0.9 \omega$. In our numerical simulations, we vary the value of $\Delta$ from $0$ (spherical harmonic trap) to $400 l_{\rm{osc}}^2=243.4 ~\mu \rm{m^2}$ (bubble trap). For $\Delta=243.4 ~\mu \rm{m^2}$ and $\Omega=0.9 \omega$, we observe that the angular momentum per particle reaches $276 \hbar$ \cite{pandey_2019_hypersonic}. Further, due to the attractive DDI, the condensate's chemical potential decreases, resulting in a decrease in sound velocity $v_s=\sqrt{\mu/m}$. Consequently, for $\Delta=21.9 ~\mu \rm{m^2}$ and $\Omega=0.9 \omega$, the local fluid velocity $v=\sqrt{\Delta_{\rm {max}}}\Omega\approx2.646~\rm{mm/s}$ significantly surpasses the sound velocity $v_s\approx 0.116~\rm{mm/s}$ \footnote{ To assess the stability of the flow at this hypersonic speed, we perturb the system by an impurity. Notably, the flow remains stable against these perturbations, as illustrated in the Appendix \ref{appA2}.}, indicating that the condensate rotates at a hypersonic speed with a Mach number $v/v_s\approx22$ \cite{guo_2020_supersonic}. Interestingly, as we increase $\Delta$, the transverse trapping frequency also rises (see Eq. (\ref{bubble}) and Fig. \ref{fig:fig1}), causing the rotation-induced droplets to merge with nearby ones and form macrodroplets immersed within a background superfluid [Fig. \ref{fig:fig5}(b)]. Further increasing the value of $\Delta$ results in the suppression of the density of the background superfluid, leading to the formation of an isolated macrodroplet crystal state [Fig. \ref{fig:fig5}(c)] along with an increase of the Mach number. For $\Delta=243.4 ~\mu \rm{m^2}$ and $\Omega=0.9 \omega$, the Mach number is close to $71$.
\label{sec4}
\section{Topological transition}
So far we have reported several intriguing phenomena that can emerge as a result of the unique bubble-shaped topology. In order to delve deeper into the role of the topology of the trap, we thoroughly investigate the impact of dynamical transitioning from a bubble-trapped BEC to a spherical-filled BEC and vice versa by tuning experimentally controlled parameter $\Delta$ with fixed $\Theta=0.0441l_{\rm{osc}}^2$.\par
\begin{figure}[t]
    \includegraphics[width=0.48\textwidth]{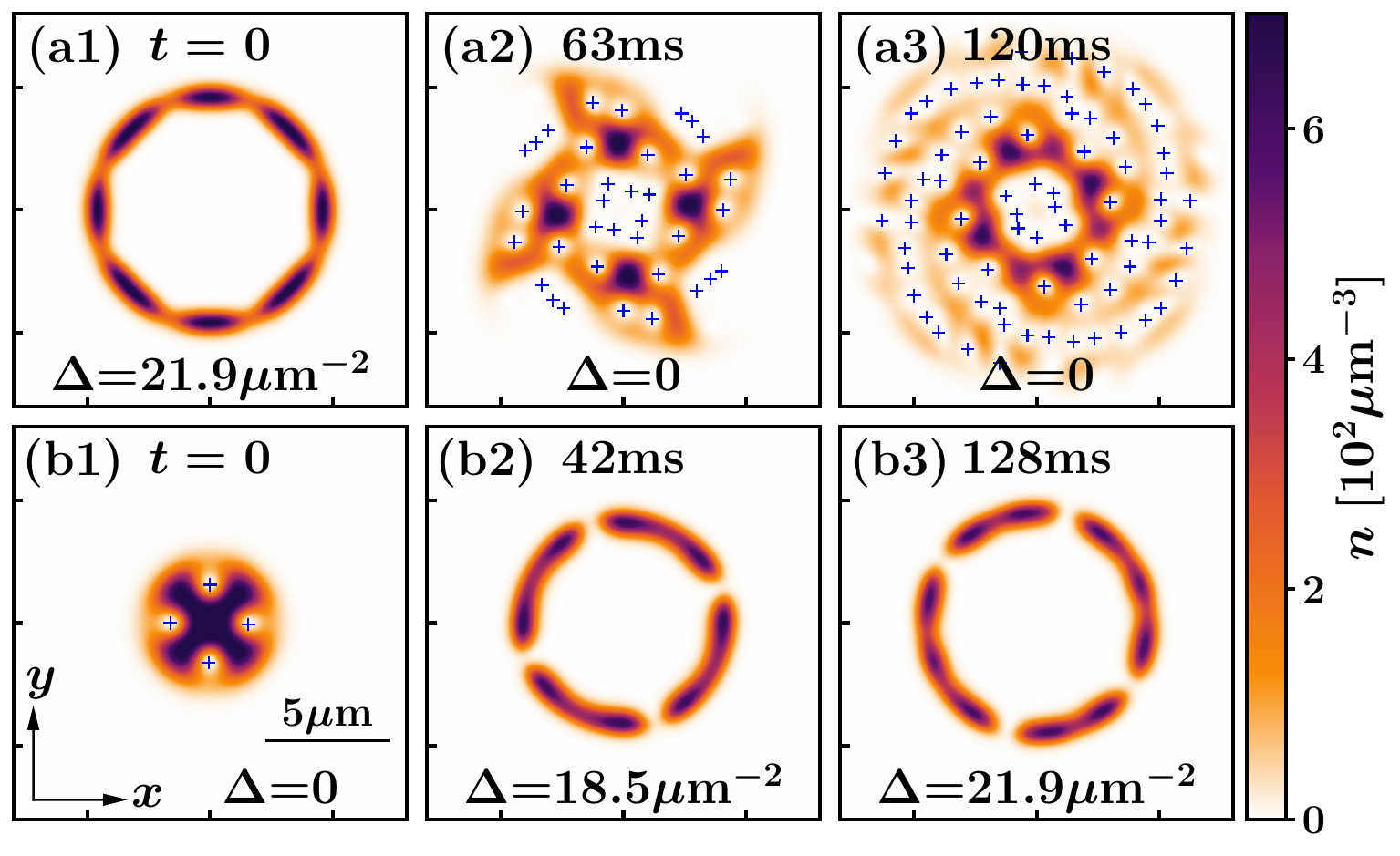}
    \caption{Snapshots of density distribution following topological transitions in the $z=0$ plane and with rotation frequency $\Omega=0.53\omega$. In the upper panel ($\rm{a}1$)-($\rm{a}3$) corresponds to the configurations during the topological change from bubble-trapped BEC to filled spherical BEC, while in the lower panel ($\rm{b}1$)-($\rm{b}3$) corresponds to the configurations during the hollowing transition from filled spherical BEC to bubble-trapped BEC. The ``+" symbol represents the presence of a vortex. A full movie of the topological transition is available in Supplementary Material.}
    \label{fig:fig6}
\end{figure}

To initiate this study, we first consider a rotation-induced supersolid state in a bubble trap with $\Omega=0.53\omega$ and $\Delta=21.9~\mu\rm{m}^2$. Starting from this ground state configuration, we linearly decrease \footnote{Irrespective of the initial state and trap rotation frequency, a sudden transition from a bubble trap to a spherical harmonic trap results in the emergence of matter-wave interference and reduction in the condensate's lifetime. See Fig. S4 in the Supplementary Material.} the parameter $\Delta$ to a final value of $\Delta=0$, representing a spherical harmonic trap. This modulation of $\Delta$ is executed over a time period of $50~\rm{ms}$,  after which we let the condensate evolve for a certain time, keeping $\Omega$ constant. Initially, owing to the bubble-shaped trap geometry and influenced by trap rotation at $\Omega=0.53\omega$, we observe that the condensate exhibits a supersolid state [Fig. \ref{fig:fig6}(a1)] and is circulating about the rotation axis of the bubble trap. This results in a relatively high angular momentum ($\langle\hat{L}_z\rangle/N\approx18\hbar$) compared to the ground state rotating with the same rotation frequency $\Omega=0.53\omega$ in a harmonic trap ($\langle\hat{L}_z\rangle/N\approx1.54\hbar$) [Fig. \ref{fig:fig6}(b1)]. During the linear decrease in the value of $\Delta$, the rotational symmetry ensures $[\hat{H}, \hat{L}_z]=0$, implying that the expectation value of $\hat{L}_z$ remains conserved. Consequently, the angular momentum of the condensate, stemming from its rotation about the rotation axis in the bubble trap, is transferred to the condensate within the spherical harmonic trap in the form of topological excitations of vortices.  Due to the significant angular momentum, the condensate's radius expands and accommodates a large number of vortices distributed in a spiral pattern [see Figs. \ref{fig:fig6}(a2) and \ref{fig:fig6}(a3)].\par
As we stated, achieving a strongly correlated quantum Hall regime necessitates a large angular momentum, surpassing the deconfinement frequency limit for harmonically trapped ultracold gases. However, instead of following a direct dynamic stirring path, one can follow the above-mentioned protocol. Starting from a rotating bubble trap with large $\Omega$ and $\Delta$, dynamically reduce the parameter $\Delta$ to zero. Given the ability to attain substantial angular momentum within the bubble trap and the conservation of angular momentum, this approach has the potential to facilitate the realization of the desired quantum Hall regime for a harmonically trapped BEC.\par
We carry out the same dynamics in the opposite direction by changing the topology from a filled sphere ($\Delta=0$) to a hollow shell ($\Delta=21.9~\mu\rm{m}^2$). Initially, with the specified parameter value of $\epsilon_{\rm{dd}}=1.31$, the condensate within the harmonic trap exhibits a superfluid state. Under the influence of trap rotation ($\Omega=0.53\omega$), the condensate demonstrates four off-centered vortices at $t=0~\rm{ms}$ [see Fig. \ref{fig:fig6}(b1)], a characteristic feature of the superfluid phase. As we increase the value of $\Delta$, the density at the center diminishes and the vortices become nested at the center within a zero-density region. As a consequence of the conservation of angular momentum, the angular momentum of the vortices transfers to the condensate, and the solid-body response of the density-modulated condensate causes its circulating motion about the rotation axis. Furthermore, analogous to the ground state of a bubble trap rotating with the same frequency ($\Omega=0.53\omega$) [see Fig. \ref{fig:fig3}(d)], the increase in local DDI induces density modulation in the condensate, leading to the formation of a supersolid state [see Figs. \ref{fig:fig6}(b2) and \ref{fig:fig6}(b3)].
\label{sec5}
\section{Conclusions and Outlook}
We have systematically examined the ground state phases of a dipolar BEC by varying the parameter $\epsilon_{\rm dd}$ within a bubble trap. Our investigation reveals that rotation applied to the bubble trap with certain ranges of rotation frequencies induces a transition from superfluid to supersolid phase, achieved through the modification of the local DDI strength. This transition to a supersolid state induced by rotation manifests as a first-order phase transition, while the transition resulting from tuning the scattering length exhibits a second-order nature. Additionally, we have observed that condensate confined in a bubble trap can sustain higher circulation for an extended duration. Notably, our findings also demonstrate the hypersonic flow, and the emergence of macrodroplets rotating with even higher Mach number at higher trap rotation frequencies and larger values of $\Delta$. Furthermore, our analysis of the hollowing transition in the rotating frame, confirms that the condensate undergoes modulation due to the confinement imposed by the bubble trap. Another aspect of our study explores the effect of topological change from a bubble trap to a filled BEC, demonstrating that tuning the trap parameter $\Delta$, can serve as a protocol to generate higher angular momentum states in harmonic confinement. This procedure has the potential to bring the interacting atomic gas into the desired quantum Hall regime \cite{sinha_2005_two_dimensional, mukherjee_2022_crystallization}. The persistent current and hypersonic velocity exhibited by atomic dipolar BECs in a rotating bubble trap could pave the way for their utilization in quantum transport applications within atomtronics \cite{amico_2022_colloquium, amico_2017_focus, ryu_2020_quantum}.\par

Regarding the non-rotating case, we recently became aware of a very recent work \cite{sanchezbaena_2024_ring} discussing the ground state phases of a dipolar BEC in a bubble trap.
\label{sec6}
\section*{Acknowledgments}
We thank Vito W. Scarola for carefully reading the manuscript and insightful discussions, and Uwe R. Fischer for their valuable comments and suggestions. We also thank Arpana Saboo for helpful discussions and carefully reading the manuscript. We acknowledge the National Supercomputing Mission for providing computing resources of PARAM Shakti at IIT Kharagpur. H. S. Ghosh gratefully acknowledges the support from the Prime Minister's Research Fellowship (PMRF), India. S. Halder acknowledges the MHRD Govt. of India for the research fellowship. S. Das acknowledges support from AFOSR FA9550-23-1-0034.
\appendix

\section{IMPACT OF ROTATION ON SUPERSOLID PHASE}\label{appA1}
\begin{figure}[H]
    \includegraphics[width=0.48\textwidth]{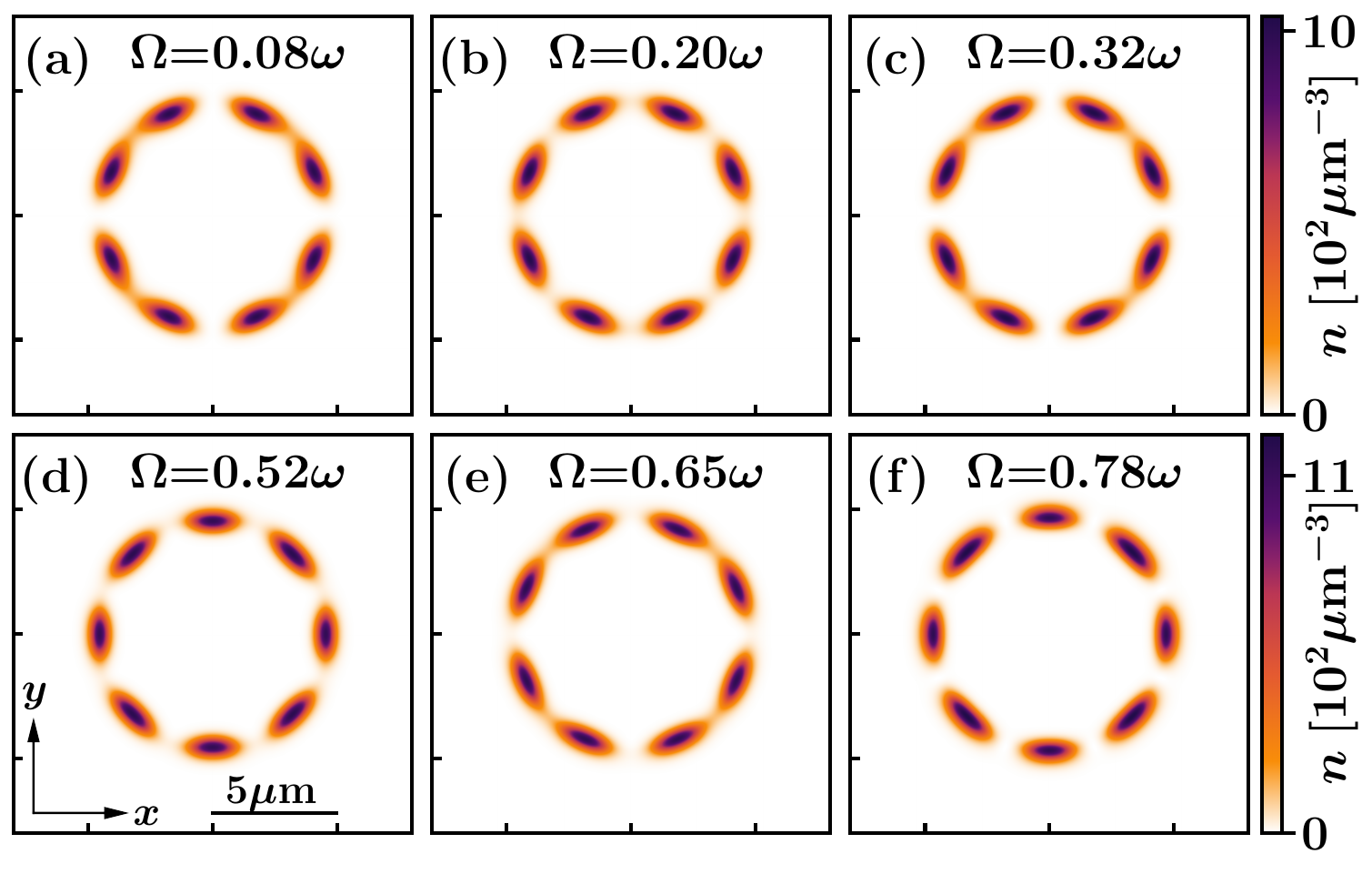}
    \caption{(a)-(f) Show the rotating ground state density profiles with $\epsilon_{\rm{dd}}=1.37$ for $\Omega=0.08\omega, 0.20\omega,0.32\omega,0.52\omega,0.65\omega,$ and $0.78\omega$ in the $z=0$ plane, respectively. Other parameters are same as of Fig. \ref{fig:fig2}(b).}
    \label{fig:fig7}
\end{figure}

\begin{figure}[b]
    \centering
    \includegraphics[width=0.48\textwidth]{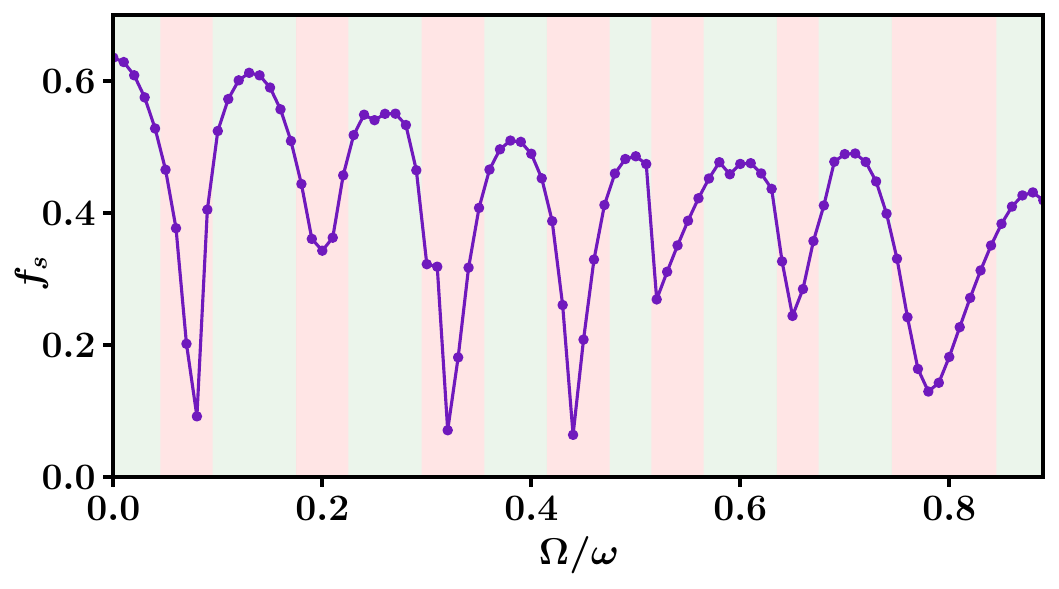}
    \caption{Superfluid fraction $f_s$ for rotating ground states at $\epsilon_{\rm{dd}}=1.37$ with trap rotation frequency $\Omega$. The dips at specific ranges of $\Omega$ values indicate the modification of the superfluid fraction. Other parameters are same as of Fig. \ref{fig:fig2}(b).}
    \label{fig:fig8}
\end{figure}

In the main text, we have demonstrated the impact of rotation on the superfluid phase of a dipolar BEC confined in a bubble trap geometry. In this addendum, we explore the effect of rotation on the supersolid state. For this investigation, we begin with a non-rotating dipolar condensate confined in a bubble trap with $\Delta=21.9~\mu \rm{m}^2$, $\Theta=0.0441 l_{\rm{osc}^2}$, and $\epsilon_{\rm{dd}}=1.37$. The ground state exhibits a supersolid state with regularly arranged ellipsoidal droplets along the trap and immersed in a background of low-density condensate with the superfluid fraction $f_s\approx 0.64$ as illustrated in Fig. \ref{fig:fig2}(b). However, as we rotate the system, the density of background superfluid showcases sudden dips in its value at certain ranges of $\Omega$ values [see Figs. \ref{fig:fig7}(a)-\ref{fig:fig7}(f)]. Consequently, the value of the superfluid fraction of the condensate drops at those
    {\unskip\parfillskip 0.0pt\par}

\noindent   specific ranges of $\Omega$ values [see Fig. \ref{fig:fig8}]. This indicates that, under the influence of trap rotation, the supersolid state transitions towards an isolated droplet phase at certain ranges of rotation frequencies. Note that, in the droplet phase ($\epsilon_{\rm{dd}}>1.45$), the system only exhibits a solid-body response with the trap rotation, and the angular momentum changes linearly with $\Omega$.

\section{STABILITY OF THE FLOW}\label{appA2}
\begin{figure}[t]
    \centering
    \includegraphics[width=0.48\textwidth]{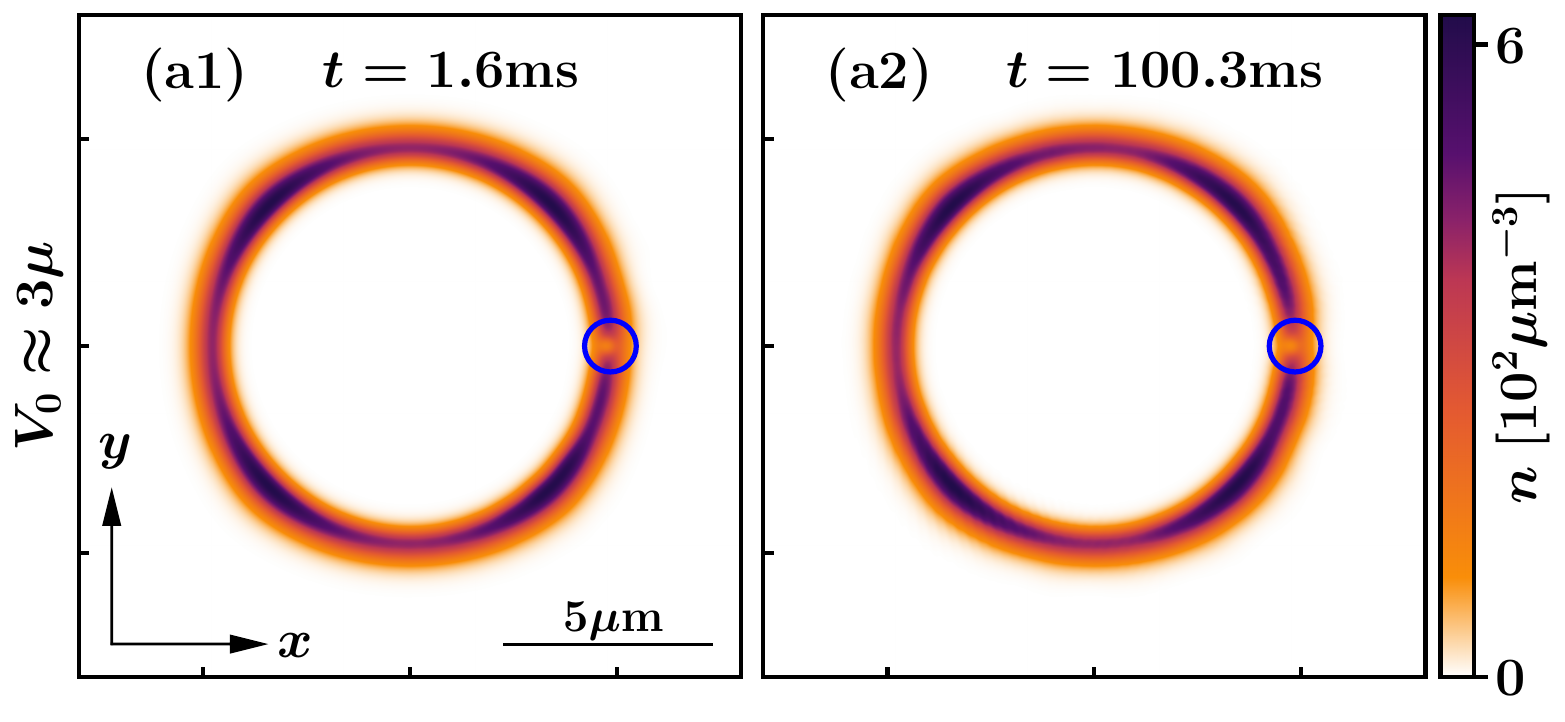}
    \caption{Snapshots of the density distribution of the rotating ground state with $\epsilon_{\rm{dd}}=1.31$, $\Delta=21.9~\mu\rm{m}^2$ and $\Omega=0.9\omega$ in the $z=0$ plane, that is perturbed by a repulsive Gaussian potential with $V_0\approx 3\mu$, $\sigma=0.2l_{\rm{osc}}$, and $x_0=4.76 ~\mu\rm{m}$, $y_0=0$, $z_0=0$. The blue circular marker in (a1) and (a2) marks the position of the impurity.}
    \label{fig:figS4}
\end{figure}

We have observed that the condensate flows at a hypersonic speed in a rotating bubble trap with a large rotation frequency. For $\Delta=21.9 ~\mu \rm{m^2}$ and $\Omega=0.9 \omega$, the condensate flows with a Mach number $v/v_s\approx22$. Typically, according to the Landau criterion, above a critical velocity, excitation emerges in the system, leading to the breakdown of the frictionless flow. However, several studies have reported that when the kinetic energy of the system becomes sufficiently high to dominate all other energy scales, it suppresses excitations even above the Landau critical velocity \cite{law_2000_motional, pavloff_2002_breakdown, paris_2017_superfluid, bradley_2016_breaking, dries_2010_dissipative, kasamatsu_2002_giant}. To illustrate the effect of perturbations on the stability of the flow, during time evolution we perturb the rotating ground state with $\epsilon_{\rm{dd}}=1.31$, $\Delta=21.9~\mu\rm{m}^2$ and $\Omega=0.9\omega$ by introducing an impurity potential of Gaussian form, defined as $V_{\rm{perturb}}=V_0\exp[-\{(x-x_0)^2 + (y-y_0)^2 + (z-z_0)^2\}/\sigma^2]$, where $V_0$ is the potential height, $\sigma$ determines the width of the potential, and $\{x_0,y_0,z_0\}$ is the impurity position. We examine the stability of the flow against perturbations for various potential heights $V_0\approx \mu,~2\mu$ and $3\mu$, where $\mu$ is the chemical potential of the system. In each case, due to hypersonic velocity, the perturbations hardly affect the flow. Thus, the flow in our system is stabilized by its own angular momentum. Fig. \ref{fig:figS4} shows the time evolution of the perturbed system with $V_0\approx 3\mu$ and $\sigma=0.2l_{\rm{osc}}$.

\bibliographystyle{apsrev4-2}
\bibliography{reference.bib}
\foreach \x in {1,2}
    {
        \clearpage
        

\section*{Supplementary material}

# Supplemental Material: Induced supersolidity and hypersonic flow of a dipolar Bose-Einstein Condensate in a rotating bubble trap

Hari Sadhan Ghosh,[1,*] Soumyadeep Halder,[1,†] Subrata Das,[1,2,‡] and Sonjoy Majumder[1,§]

[1]*Department of Physics, Indian Institute of Technology Kharagpur, Kharagpur 721302, India*
[2]*Department of Physics, Virginia Tech, Blacksburg, Virginia 24061, USA*
(Dated: August 18, 2024)

## I. NON-ROTATING GROUND STATE

In a static bubble trap, we observe different ground state phases of a dipolar BEC depending on the value of $\epsilon_{\text{dd}}$. Here, we demonstrate the three-dimensional density distributions of those phases in Fig. S1. For $\epsilon_{\text{dd}} < 1.35$, the ground states are in the superfluid phase [see Figs. S1(a1) and S1(a2)]. As $\epsilon_{\text{dd}}$ is increased, beyond a critical threshold ($\epsilon_{\text{dd}} \approx 1.35$), the DDI begins to dominate over the contact interaction, leading to modulation in the density profile. In the interval $1.35 < \epsilon_{\text{dd}} < 1.43$, the condensate exhibits a periodic arrangement of ellipsoidal droplets, elongated along the polarisation direction, with a background superfluid density [see Figs. S1(b1) and S1(b2)]. With further increase in the value of $\epsilon_{\text{dd}}$, background superfluid density diminishes, and an incoherent crystal of isolated droplets is formed [see Figs. S1(c1) and S1(c2)].

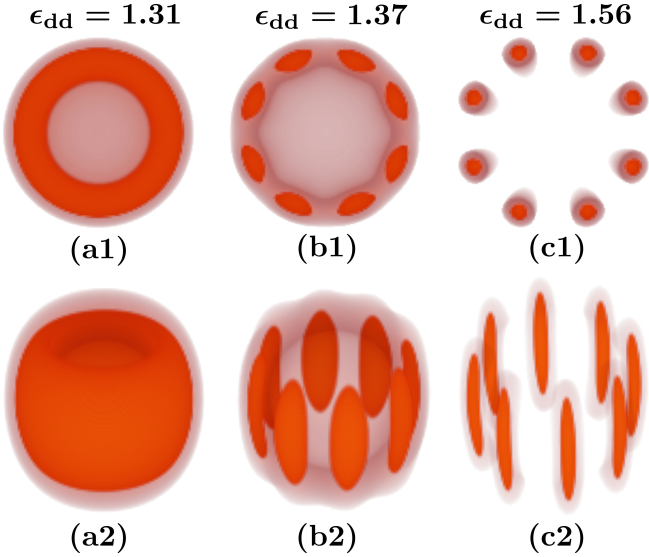

FIG. S1. Three-dimensional isosurface density distribution of the ground states for different values of $\epsilon_{\text{dd}}$. Three different phases are superfluid phase ((a1),(a2)), supersolid phase ((b1),(b2)) and droplet phase((c1),(c2)). The upper panel is the top view of the density profiles, while the lower panel is the side view.

---
[*] harisadhanghosh4@gmail.com
[†] soumya.hhs@gmail.com
[‡] subrata@vt.edu
[§] sonjoym@phy.iitkgp.ac.in

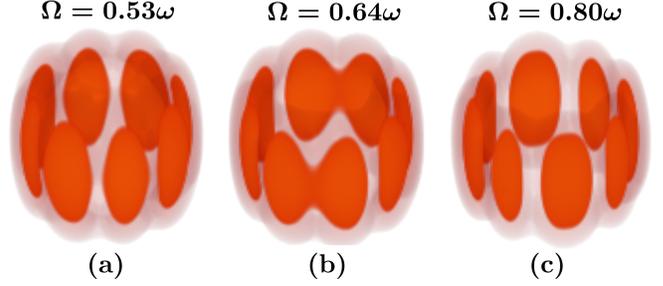

FIG. S2. Three-dimensional isosurface density distribution of the rotating ground sates with $\epsilon_{dd} = 1.31$ and the trap rotation frequencies $\Omega = 0.53\omega, 0.64\omega$, and $0.80\omega$.

## II. ROTATING GROUND STATE

The density profiles in the $z = 0$ plane of the rotating ground states are illustrated in the main text for specific rotation frequencies and $\epsilon_{dd} = 1.31$. Here, we have depicted the three-dimensional isosurface density profiles in Fig. S2 for better visualization. With zero rotation frequency and for $\epsilon_{dd} = 1.31$, the ground state exhibits superfluid phase [see Figs. S1(a1) and S1(a2)]. As we rotate the system, the condensate undergoes a transition from superfluid to supersolid phase at certain ranges of $\Omega$ values [see Fig. S2]. Note that, both the non-rotating and rotation-induced supersolid states are not stabilized within the mean-field framework. Therefore, Lee-Huang-Yang (LHY) correction is required for stabilization.

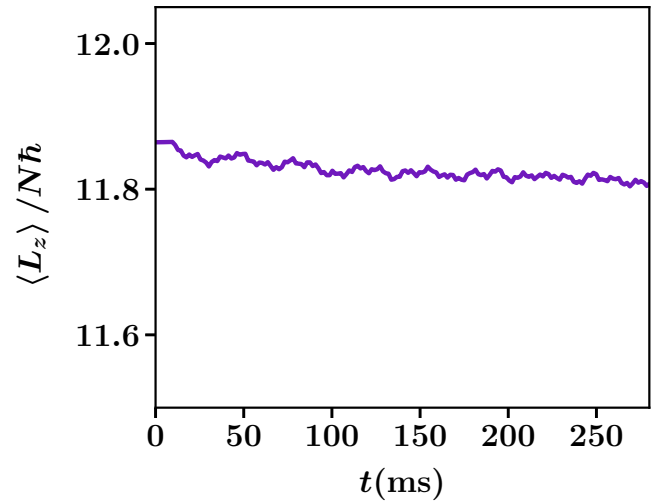

FIG. S3. Variation of angular momentum $\langle \hat{L}_z \rangle / N\hbar$ of the condensate with time after stopping the rotation of the trap.

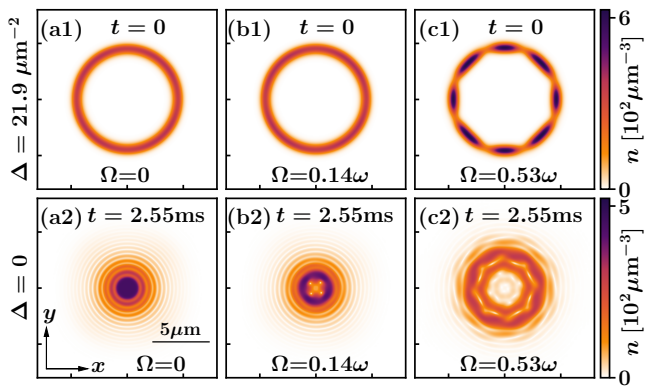

FIG. S4. The emergence of self-interference pattern due to sudden transition from bubble trap to harmonic trap. (a1), (b1), and (c1) represent a non-rotating superfluid state, a rotating superfluid state with $\Omega = 0.14\omega$ and a rotation-induced supersolid state with $\Omega = 0.53\omega$ at $t = 0$, respectively. (a2),(b2), and (c2) are the corresponding density distribution at $t = 2.55$ ms after sudden switching into a harmonic trap from a bubble trap.

In the main text, we have demonstrated that the Dy condensate within rotating bubble trap geometry spins about the rotation axis and possesses high angular momentum. We also observe that this flow sustains for a long period of time even after halting the rotation of the trap. Following the cessation of trap rotation, in Fig. S3 we have shown the variation of the expectation value of angular momentum per particle of the initial ground state, generated at $\Omega = 0.44\omega, \epsilon_{\rm dd} = 1.31$ and $\Delta = 21.9$ $\mu$m$^2$.

After stopping the rotation of the trap, the expectation value of the angular momentum remains nearly constant, with a minimal change from $11.86\hbar$ to $11.81\hbar$ occurring over a span of 286 ms.

## III. TOPOLOGICAL TRANSITION

In the main text, we have discussed the effect of the dynamical transition from bubble trap to harmonic trap by linearly decreasing the value of $\Delta$ to a final value of $\Delta = 0$ over a duration of 50 ms. Here, we showcase the impact of the sudden transition from the value of $\Delta = 21.9$ $\mu$m$^2$ (bubble trap) to $\Delta = 0$ (harmonic trap) on the trapped BEC. Upon quenching the value of $\Delta$ to 0, the inner radius of the BEC rapidly moves towards the center. As a result, the condensate superimposes with itself at the trap center, leading to the manifestation of alternating circular self-interference patterns. To initiate this study, we consider three different ground states with $\epsilon_{\rm dd} = 1.31$ and $\Delta = 21.9$ $\mu$m$^2$ (a) a non-rotating superfluid state [Fig. S4(a1)], (b) a rotating superfluid state with $\Omega = 0.14\omega$ [Fig. S4(b1)] and (c) a rotation-induced supersolid state with $\Omega = 0.53\omega$ [Fig. S4(c1)]. In each of these three distinct cases, when we abruptly change the value of $\Delta$ to 0, we observe the emergence of circular interference patterns after a very short period of time [see Figs. S4(a2), S4(b2) and S4(c2)]. Zobay and Garraway also observed similar interference phenomena [1] when they quickly changed the shell-shaped trap to a harmonic trap. To avoid the self-interference phenomenon during the topological transition we tune the value of $\Delta$ with a slow linear ramp.


    }

\end{document}